# Effective modeling of ground penetrating radar in fractured media using analytic solutions for propagation, thin-bed interaction and dipolar scattering

Alexis Shakas[1*] and Niklas Linde[1]

[1]Applied and Environmental Geophysics Group, Institute of Earth Sciences, University of Lausanne, Switzerland.

* Corresponding author: Alexis Shakas
University of Lausanne
Géopolis - bureau 3321
CH-1015 Lausanne
Email :        Alexis.Shakas@unil.ch
Phone :        +41 21 692 43 06
Fax :          +41 21 692 44 05

# ABSTRACT

We propose a new approach to model ground penetrating radar signals that propagate through a homogeneous and isotropic medium, and are scattered at thin planar fractures of arbitrary dip, azimuth, thickness and material filling. We use analytical expressions for the Maxwell equations in a homogeneous space to describe the propagation of the signal in the rock matrix, and account for frequency-dependent dispersion and attenuation through the empirical Jonscher formulation. We discretize fractures into elements that are linearly polarized by the incoming electric field that arrives from the source to each element, locally, as a plane wave. To model the effective source wavelet we use a generalized Gamma distribution to define the antenna dipole moment. We combine microscopic and macroscopic Maxwell's equations to derive an analytic expression for the response of each element, which describe the full electric dipole radiation patterns along with effective reflection coefficients of thin layers. Our results compare favorably with finite-difference time-domain modeling in the case of constant electrical parameters of the rock-matrix and fracture filling. Compared with traditional finite-difference time-domain modeling, the proposed approach is faster and more flexible in terms of fracture orientations. A comparison with published laboratory results suggests that the modeling approach can reproduce the main characteristics of the reflected wavelet.

## 1. Introduction

Identification and characterization of permeable fractures within rock formations is a central research topic in hydrology (National Research Council, 1996). The flow and transport behavior in fractured media can be very complex and difficult to infer from traditional hydrological experiments (Neuman, 2005). A promising approach is to combine hydrologic measurements with ground penetrating radar (GPR) data (e.g. Olsson et al., 1992). Both surface reflection and cross-borehole tomographic monitoring studies have been used to infer the spatial distribution of tracer plumes and to dynamically image tracer transport through preferential flow paths (Birken and Versteeg, 2000; Tsoflias et al., 2001; Day-Lewis et al., 2003; Talley et al., 2005; Becker and Tsoflias, 2010; Dorn et al., 2011, 2012a). Furthermore, the ability of GPR to provide information about mm-thick fractures has been demonstrated theoretically (Hollender and Tillard, 1998; Bradford and Deeds, 2006; Tsoflias and Hoch, 2006), through controlled experiments (Grégoire and Hollender, 2004; Tsoflias et al., 2004; Sambuelli and Calzoni, 2010) and by field-based investigations (Tsoflias and Hoch, 2006; Sassen and Everett, 2009; Dorn et al., 2011, 2012b). In the complex environment found in most fractured rock systems, efficient and effective interpretation of GPR measurements must rely on forward models that accurately simulate the experiments.

When an electromagnetic wave impinges on a thin layer, a series of complex interference phenomena occur that alter both the phase and amplitude of the reflected and transmitted waves. Such phenomena have been studied extensively in optics and exact solutions are available by applying the macroscopic Maxwell's equations and associated boundary conditions on the sides of a dielectric slab (e.g., a fluid filled fracture). These solutions have been used in geophysics to describe how the GPR signal reflected from fractures varies as a function of material properties, fracture thickness (aperture) and orientation (Tsoflias and Hoch, 2006; Tsoflias and Becker, 2008).

Numerical GPR forward modeling schemes do not incorporate the analytic nature of the effective reflection coefficients since space discretization and medium parameterization implicitly account for boundaries, across which the macroscopic Maxwell's equations are solved. As spatial discretization becomes finer, the macroscopic numerical solutions approach the analytically derived Fresnel reflection and transmission coefficients. However, the finite spatial discretization gives rise to errors, especially when modeling irregular geometries or fine-scale structures. Sub-discretization schemes have been recently proposed (e.g., Diamanti and Giannopoulos, 2009) but the computational demand still remains for 3D implementations. Moreover, irregular geometries still pose a problem since FDTD codes usually implement a Cartesian grid and tilted planar surfaces are not discretized exactly; a known problem that is often referred to as "staircasing". In numerical solvers based in the time domain, insufficient temporal sampling can also give rise to numerical dispersion (Bergmann et al., 1998). Ray-tracing algorithms can include effective reflection coefficients, but they rely on the plane wave assumption being valid everywhere along an interface and only consider the far-field region of electromagnetic radiation. Furthermore, ray-tracing workflows are often based on algorithms developed for seismic processing (Dorn et al., 2012b) and ignore the polarized response of GPR sources and reflections.

A more general approach is to consider a fracture as a polarizable dielectric and conductive anomaly, in which many infinitesimal dipoles are induced and oscillate in response to the incident field. This approach is exactly described by the microscopic Maxwell's equations (e.g. Purcell and Smith, 1986), in which matter is seen as a collection of polarizable particles. The macroscopic boundary conditions can then be derived as limiting cases of the microscopic approach through the Ewald-Oseen extinction theorem (Fearn et al., 1996). The macroscopic approach is thus an averaged version of the microscopic formulation, the latter not only being correct in the quantum regime but also more intuitive (Feynman et al., 1969). A numerical modeling application of the microscopic Maxwell's equations has been used extensively by the astrophysical community to describe light scattering from dielectric objects — see Yurkin and Hoekstra (2008) for an overview — but we are not aware of applications to GPR scattering.

We propose a forward modeling approach that uses analytic solutions to simulate the propagation of electromagnetic waves within homogeneous media and the scattering of the waves from fractures. The fractures are seen as dielectric and conductive anomalies that are polarized by the incident EM field and are defined as rectangular planes with a given midpoint, azimuth, dip, thickness and material filling. Each fracture plane is discretized into polarizable elements, a formulation which enables simulating heterogeneous tracer concentrations in the fractures by varying the electrical properties of each element over time, and also accounts for the change in direction and magnitude of the incident electric field along the fracture plane. The elements are modeled as infinitesimal dipoles that are polarized linearly and in parallel to the incident electric field. The main difference from the astrophysical formulation is that we only assign effective dipoles along the plane of the fracture. To account for the effect of the dipoles along the direction normal to the fracture plane we apply the Ewald-Oseen extinction theorem and scale the dipoles by the effective reflection coefficients of a thin layer. Another difference is that we only consider the incident field caused by the external source and do not account for interactions between elements. We use analytical expressions of the Maxwell equations in a homogeneous space to describe the propagation of the EM field to and from each element and allow for frequency-dependent attenuation and dispersion through the Jonscher constitutive formulation (Jonscher, 1999). The resulting forward modeling scheme is free from boundary effects related to the modeled domain size and also from discretization errors. We begin by describing the theory before proceeding with how we discretize a fracture, and, finally, we compare our forward modeling scheme to simulations based on a well-established numerical code and to laboratory data.

## 2. Theory

The electromagnetic properties of matter that characterize the velocity, attenuation and dispersion of electromagnetic (EM) energy in dielectric media are the magnetic permeability $\mu$ (N A$^{-2}$), the electric permittivity $\varepsilon$ (F m$^{-1}$) and the conductivity $\sigma$ (S m$^{-1}$), or equivalently the resistivity, $\varrho$ ($\Omega$ m), with $\varrho = \sigma^{-1}$. These parameters are in general complex-valued and frequency dependent, while for many practical geophysical purposes it is safe to assume the magnetic permeability to be constant and equal to the value in vacuum, $\mu_0 = 4\pi\times10^{-7}$ N A$^{-2}$. Reflections and transmissions arise at the boundary between contrasting media and are a form of energy scattering. For geophysical purposes it is customary to use the macroscopic Maxwell equations as the governing physical principles to describe such systems (Zonge et al., 1991) and the link between the propagating field to a given medium is made through the constitutive relations, **D** = $\varepsilon$ **E** and **J** = $\sigma$ **E**, where **E** (V m$^{-1}$) is the incident electric field arising from a distant source, **J** (A m$^{-2}$) is the resulting current density and **D** (C m$^{-2}$) is the electric displacement field.

There is a theoretical distinction between permittivity and conductivity because the first describes polarization effects resulting from bound charge and the second conduction effects resulting from free charge. In practice, these two parameters can be combined since one can only measure the in-phase and out-of-phase components of the current (Hollender and Tillard, 1998). It is thus convenient to define the effective permittivity $\varepsilon_e$ (F m$^{-1}$) with real and imaginary parts that characterize the propagation properties of the material: wave velocity, attenuation and dispersion.

*2.1. The microscopic viewpoint*

While the electric displacement field **D** was introduced by Maxwell and is proportional to the "bound" charge density within a dielectric (Purcell and Smith, 1986), it is only an approximation resulting from spatial averaging of a microscopic process that involves interaction between fields and particles that make up matter. The microscopic description was introduced by Lorentz (1916) and considers a dielectric as a collection of particles that undergo electronic polarization from an externally applied electric field. The applied field exists independently of the dielectric medium and travels through the dielectric medium at the speed of light in vacuum, that is, in the free space between the particles of the dielectric. As it travels through the medium it polarizes the particles that make up the dielectric, inducing moments of charge distribution in each particle. For neutral dielectrics it is only the electric dipole moment that needs to be considered and polarization can be seen as the result of an induced charge separation that generates an electric dipole moment $p = q\mathrm{d}L\hat{r}_d$ for separation $dL$ (m) between two opposite charges of equal magnitude $q$ (C) and orientation $\hat{r}_d$. The electric field produced by such a dipole moment can be accurately calculated for an observation distance $r$ (m) much larger than the charge separation $dL$ producing the dipole moment and becomes exact in the limiting case, $\frac{\mathrm{d}L}{r} \to 0$, in which the induced dipole is often called a point dipole. The electric field of the point dipole is given by:

$$E_d(r,p) = \frac{1}{4\pi\varepsilon_0}\left\{k^2(\hat{r}\times p)\times\hat{r} + (3\hat{r}(\hat{r}\cdot p) - p)\left(\frac{1}{r^2} - \frac{\mathrm{i}k}{r}\right)\right\}\frac{e^{\mathrm{i}kr}}{r} \qquad (1)$$

where $\hat{r}$ is a unit vector pointing from the point dipole to the point of observation $r = \mathbf{r} \{\hat{r}, r$ (m) is the magnitude of $\mathbf{r}$, $\varepsilon_0 = 8.854\ldots\times 10^{-12}$ F m$^{-1}$ is the electric permittivity in vacuum and $k$ (rad m$^{-1}$) is the wavenumber in vacuum. The vacuum wavenumber is given by $k = \omega\, c^{-1}$, where $\omega$ (rad s$^{-1}$) is the angular frequency and $c = 299\ 792\ 458$ m s$^{-1}$ is the speed of light in vacuum. Equation (1) includes the near, intermediate and far-fields generated by a dipole $\mathbf{p}$ located at the origin of the coordinate system. We use the subscript d in the electric field ($\mathbf{E}_d$) to denote that it corresponds to a point dipole. A generalized expression for the electric field at an arbitrary location $\mathbf{r}$ generated from a dipole located at $\mathbf{r}'$ is easily obtained through the substitution $\mathbf{r} \rightarrow \mathbf{r} - \mathbf{r}'$. A detailed derivation of Eq. (1) can be found in classical electrodynamics textbooks (e.g., Jackson, 1998).

Each particle in a dielectric medium is polarized by a superposition of the applied field generated by a source far away and the fields generated by all the other particles present in the dielectric. For a uniform distribution of polarizable particles and at large observation distances compared to the inter-particle spacing, one can define the average electric dipole moment $\langle p \rangle$ (C m) per unit volume $V$ (m$^3$) as the polarization density $P = \langle p \rangle / V$ (C m$^{-2}$). The average dipole moment of a homogeneous and isotropic region that is polarized by an incident plane wave gives the same polarization response as one would get by summing the fields of the individual particles. The polarization density $\mathbf{P}$ links the microscopic approach to the macroscopic description through the process of spatial averaging of the dipoles. The electric displacement field can be written explicitly as $\mathbf{D} = \varepsilon_0 \mathbf{E} + \mathbf{P}$, highlighting that the macroscopic description implicitly includes the contribution from all the polarized particles in the electric displacement field, $\mathbf{D}$, through spatial averaging (Russakoff, 1970).

*2.2. From a microscopic to a macroscopic description*

The equivalence between the microscopic and macroscopic formulations has been rigorously proven through the Ewald-Oseen extinction theorem (Born and Wolf, 1999). This theorem states that for homogeneous and isotropic media that are linearly polarized by an externally applied field, the interaction between all the induced dipoles exactly cancels out parts of the applied field such that the resulting field propagates exactly as the electric displacement field $\mathbf{D}$ predicted by the macroscopic Maxwell's equations. This reduces the macroscopic theory to a special case of the microscopic approach, which is in its nature a more fundamental and intuitive description (Feynman et al., 1969).

Numerical implementations of the microscopic interactions between polarized dipoles and light have been long used by the astrophysical community. The discrete dipole approximation (DDA), or coupled-dipole approximation as introduced by Purcell and Pennypacker (1973) replaces a dielectric object with electric dipoles that are polarized by the local electric field. The local electric field takes into account the field radiated and induced by all the dipoles present, as well as the incoming field. This makes the DDA highly suitable for descriptions of irregular objects and the results compare well with exact theories, such as Mie and Rayleigh scattering (Yurkin and Hoekstra, 2007). Another important benefit of the DDA formulation compared to other numerical methods is that it does not require a bounded model domain; see Kahnert (2003) for an overview of numerical methods in EM scattering theory.

*2.3. Reflection and transmission from scattering*

The Ewald-Oseen extinction theorem applied to two semi-infinite regions of differing dielectric materials can be used to derive reflection and transmission coefficients (Fearn et al., 1996) that are identical to the Fresnel coefficients obtained by solving the macroscopic Maxwell's equations and matching boundary conditions across an interface. In the microscopic regime, the fields scattered from each particle superimpose exactly to give a resulting reflection and transmission that effectively occurs at the "boundary" between the two materials. It is worthwhile to note that the extinction of the applied field actually takes place throughout the whole volume of the dielectric, and not at the boundary as the macroscopic approach suggests. Moreover, the microscopic approach can account for conduction currents (Ballenegger and Weber, 1999).

In the case of a plane wave obliquely incident on a homogeneous dielectric layer of finite thickness (i.e. a planar fracture), the microscopic formulation successfully reduces to the effective Transverse Electric (TE) and Transverse Magnetic (TM) reflection coefficients $R_{e,\{TE,TM\}}$ that are identical to those derived in optics (Lai et al. 2002):

$$R_{e, \{TE, TM\}} = \frac{R_{\{TE, TM\}}(1-e^{-2i k_t d})}{1-R^2_{\{TE, TM\}}e^{-2i k_t d}} \tag{2}$$

where the effective reflection coefficient for each mode can be readily computed by replacing the interface reflection coefficients, for TE or TM modes, in Eq. (2). The interface reflection coefficients depend on the incidence angle $\theta$ of the incoming wave, and are given by the following two equations:

$$R_{TE} = \frac{\mu_t k_b \cos\theta - \mu_b \sqrt{k_t^2 - k_b^2 \sin^2\theta}}{\mu_t k_b \cos\theta + \mu_b \sqrt{k_t^2 - k_b^2 \sin^2\theta}} \tag{3}$$

$$R_{TM} = \frac{\mu_t k_b \sqrt{k_t^2 - k_b^2 \sin^2\theta} - \mu_b k_t^2 \cos\theta}{\mu_t k_b \sqrt{k_t^2 - k_b^2 \sin^2\theta} + \mu_b k_t^2 \cos\theta} \quad . \tag{4}$$

To compute the wavenumber $k_n$, where the subscript $n$ is used to denote a medium with effective permittivity $\varepsilon_{e,n}$ and magnetic permeability $\mu_n$, one can use the following relation: $k_n = \omega (\varepsilon_{e,n} \mu_n)^{1/2}$. In Eqns. (2-4) above, and in equations to follow, we use the subscript b to index the homogeneous matrix and t to index the dielectric and conductive thin-layer of thickness $d$. Frequency dependence of the reflection coefficient arises both through the explicit presence of the angular frequency in the wavenumber formula and the implicit frequency-dependence of the electric permittivity. These effective reflection coefficients have been successfully used in geophysical applications to describe GPR reflections from thin layers (Grégoire and Hollender, 2004; Tsoflias et al., 2004; Bradford and Deeds, 2006; Deparis and Garambois, 2008; Sassen and Everett, 2009).

### *2.4. Fractures seen as dipole scatterers*

Reflections from a fracture with a homogeneous material filling would be accurately described by the effective reflection coefficient (Eq. 2) if the incoming electromagnetic field is a plane wave over the whole extent of the fracture. Since the wavelengths are often comparable in scale to the extent of the fractures, one can hardly expect the incoming field to strike with the same angle throughout. Therefore, energy exchanges across a fracture are often inadequately recovered by the Fresnel coefficients. We propose to circumvent this problem by discretizing fractures into regions along which the plane wave assumption is approximately valid. We can then treat the discretized regions from a microscopic perspective and describe the reflected energy using a combination of effective reflection coefficients and scattering from polarized dipoles.

We consider fractures embedded within a homogeneous and isotropic background dielectric medium with effective electric permittivity $\varepsilon_{e,b}$ and magnetic permeability $\mu_b$ (e.g. fractures in a uniform rock matrix). We account for the interaction of the electric field with this medium by replacing the vacuum wavenumber $k$ in Eq. (1) by the wavenumber calculated in the background medium, $k_b = \omega\, (\varepsilon_{e,b}\, \mu_b)^{1/2}$. We thus replace the dipole-generated electric field ($\mathbf{E}_d$) with the electric displacement field ($\mathbf{D}_d$) by making the following substitution to Eq. (1):

$$k \to k_b \Rightarrow E_d(x,y,z,p) \to D_d(x,y,z,p). \tag{5}$$

We proceed by discretizing each fracture into regions of constant thickness and length with homogeneous and isotropic electrical properties, which we refer to as elements. The homogeneity of each element allows us to assume a continuous distribution of dipoles within its volume with a response that can be described by the polarization density $\mathbf{P}$. Furthermore, by imposing the length of each element to be small compared to the wavelength of the incoming electric field, we can approximate the incoming wave to be plane over the extent of an element.

The electric displacement field ($\mathbf{D}^m$) scattered from an element indexed $m$ is the superposition (summation) of the fields generated by all the dipoles within the element:

$$D^m(x,y,z) = \sum_{n=1}^{N} D_d^m(x,y,z,\langle p_n^m \rangle) = D_d^m(x,y,z,\oiiint P^m(x',y',z')\, dx'\, dy'\, dz'). \tag{6}$$

The summation in Eq. 6 is taken over the individual dipoles of a homogeneous element, but for observation distances that are large compared to the inter-dipole spacing we can replace this summation with a volume integral of the polarization density $\mathbf{P}$. We use primed coordinates to integrate over the volume of the element and unprimed coordinates to denote a different coordinate system outside the element that we call the experimental coordinate system, in which the observed field is measured. The element coordinate system is related to the experimental coordinate system by a dip and azimuth and the components of the electric field, when moving from one coordinate system to the other, can be calculated without loss of accuracy.

The Ewald-Oseen extinction theorem can be applied to reduce the dipole interaction along the direction normal to the element, $z'$, to a process that is effective at the first intercepting boundary of the element with the incoming field. We denote the location of the first intercepting boundary by $z_c$. The reduction is achieved by the following substitution:

$$\iiint_{x',y',z'} P^m(x', y', z') \, dx' \, dy' \, dz' \rightarrow \iint_{x',y'} R_e^m P^m(x', y', z_c) \, dx' \, dy', \tag{7}$$

where the polarization density **P** is replaced by a surface polarization density, $P$ (C m$^{-1}$). We use a round hat over vector quantities to indicate that they are surface variables that are evaluated over the first intersecting boundary of an element.

Furthermore, by allowing only elements with side length much smaller than the wavelength of the incoming wave, we can assume the incoming wave to be plane over the extent of an element. We use this assumption to reduce the surface integral over the intersecting boundary of the element to a simple multiplication by the boundary's area ($A^m$). Since the incoming wave is assumed to be a plane over the extent of an element, we only need to compute the polarization strength at one location along the first intercepting boundary and choose the center of the boundary, which we denote by ($x_c, y_c, z_c$). The surface integral of Eq. (7) then becomes:

$$\iint_{x',y'} R_e^m P^m(x', y', z_c) \, dx' \, dy' = A^m \, R_e^m P^m(x_c, y_c, z_c). \tag{8}$$

The right hand side term of Eq. (8) has the units of a dipole moment, but accounts for the total dipolar strength of an element. We call this term an "effective dipole" and denote it with the symbol $p_e^m$ (C m) for an element indexed $m$.

The effective dipole is proportional to the incoming electric field arriving from the antenna source, that we model as a dipole **p**$_a$ (C m) fixed at an arbitrary position ($x_a, y_a, z_a$) in the experimental coordinate system (more on this in section 3.2). We denote the electric displacement field arriving from the antenna source by $D_d$(C m$^{-1}$) to indicate that this field is computed along the first intersecting boundary of an element. The effective dipole is then given by:

$$p_e^m = A^m \, R_e^m P^m(x_c, y_c, z_c) = A^m \, R_e^m D_d\big(x_c, y_c, z_c, p_a(x_a, y_a, z_a)\big). \qquad (9)$$

The field $D_d$ in Eq. (8) can be decomposed into a TE and TM mode and the corresponding effective reflection coefficients, Eqs. (2-4), can be used.

In summary, we use a microscopic approach up-scaled to the size of a dipole element that acts as an effective dipole, polarized by the source dipole (antenna), modulated by its effective reflection coefficient and scaled by its area of intersection with the incoming field. For a collection of many elements, the total scattered field is a superposition of the fields from every effective dipole. For M dipole elements, the total measured field **D** is:

$$\begin{aligned} D(x, y, z) &= \sum_{m=1}^{M} D^m(x, y, z) = \sum_{m=1}^{M} D_d^m(x, y, z, p_e^m) \\ &\sum_{m=1}^{M} D_d^m\left(x, y, z, A^m \, R_e^m D_d^m(x_c, y_c, z_c, p_a(x_a, y_a, z_a))\right) \end{aligned} \qquad (10)$$

The recursive use of Eq. (1) in Eq. (10) arises from the representation of the antenna source and the fracture element as dipoles and highlights the similarity of our approach to the discrete dipole approximation used in astrophysics. A schematic of this interaction is shown in Fig. 1.

## 3. Implementation

We now propose a strategy to implement the aforementioned effective-dipole forward modeling scheme. We begin by stating the underlying assumptions, we then proceed to give an effective description of the emitted GPR signals, proceed to generalize the method so that frequency-dependent properties can be assigned to the materials, and finally examine how to effectively discretize a fracture for a given survey configuration.

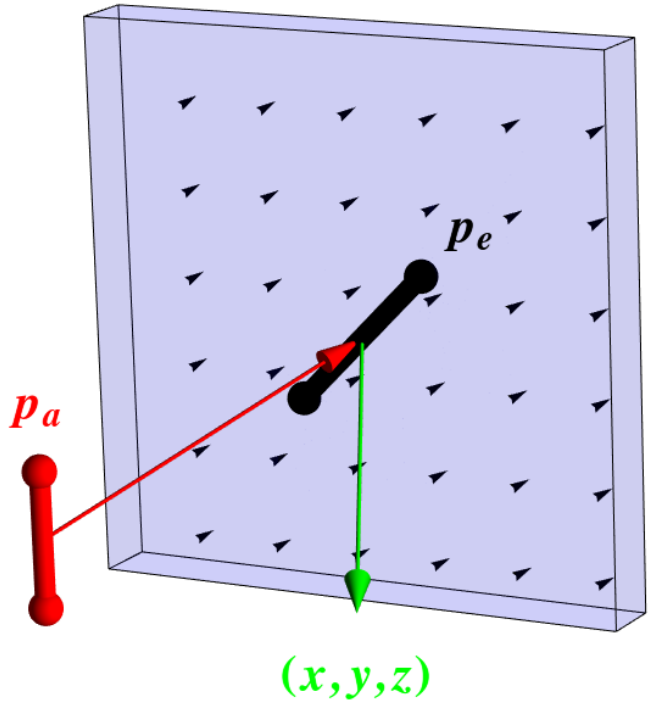

**Fig. 1.** Schematic of an element. The direction of the incident plane wave is shown by the (red) arrow pointing from the source dipole **p**$_a$ to the induced effective dipole **p**$_e$. The secondary field (green) generated by the element is measured at the receiver location (*x,y,z*). The smaller (black) arrows are the "individual" dipoles within the dielectric.

*3.1. Assumptions about subsurface properties*

The validity of the presented forward modeling scheme is limited by the following assumptions about the subsurface:

1. The rock matrix is a homogeneous and isotropic dielectric medium. This assumption is reasonable for many applications in fractured rock, for which the dominant heterogeneities are related to the fractures;
2. The fracture is a planar rectangular surface (other geometries could easily be considered);
3. The fracture filling is isotropic and polarizes linearly by the incoming field. The linear response is valid since the emitted signal is weak and well within the linear limits of dielectric Earth materials;

4. Magnetic effects are not taken into account and the magnetic permeability is set equal to $\mu_0$ everywhere;

The forward problem is solved in the frequency domain.

*3.2. Source signal generation*

In a resistively loaded dipole antenna, a short pulse of current is exponentially damped along the antenna length and produces a dominant dipole moment that induces an electric field. The resulting electric field has a complicated structure arising from asymmetries in the antenna design, antenna coupling with the surrounding medium, noise in the system and other (often unknown) sources of error. We partly account for such uncertainties by modeling the antenna dipole moment with a generalized Gamma distribution (Stacy, 1962), thus allowing much more flexibility in the resulting shape of the dipole moment compared with a Gaussian distribution. The frequency-dependent response of the antenna dipole moment is given by:

$$p_a(\omega, \alpha, \beta, \gamma, \mu) = K\,(\omega - \mu)^{\alpha\gamma - 1} \mathrm{Exp}\left[-\left(\frac{\omega - \mu}{\beta}\right)^\gamma\right]\hat{r}\,;\ \omega > \mu \qquad (11)$$

where $\alpha$, $\beta$ and $\gamma$ are positive parameters that control the shape of the dipole moment distribution, $\hat{r}$ is the orientation of the antenna, $\mu$ is a location parameter below which the distribution is zero-valued and K (C m) is a normalization constant. Equation (11) reduces to a Gaussian distribution for $\alpha = 0.5$, $\gamma = 2$, $\mu = 0$ and with standard deviation $\beta$. Our approach to model an effective antenna source is similar, but not equivalent, to the newly-established full-waveform inversion method (Ernst et al., 2007). The latter approach has been succesfully used in several studies to invert GPR data where the source wavelet is unknown (e.g. Klotzsche et al., 2013).

*3.3. Frequency-dependent polarization*

The polarization described until now is a form of electronic polarization, in which the bound charge distribution of a particle is instantaneously "reshaped" by the incoming field. The amount of reshaping is quantified by the electric susceptibility, $\chi = \varepsilon - 1$, and is in general frequency-dependent. The restoration time accompanied with the bound charge moving back in place causes an out-of-phase response that leads to energy loss. The overall effect of electromagnetic waves propagating through dielectric matter is to generate both polarization and conduction currents, a process often referred to as dielectric relaxation (Jonscher, 1999). For low-loss dielectric media, the Jonscher 'universal dielectric response' effectively describes the effects of polarization and static conduction for typical GPR frequencies (Hollender and Tillard, 1998). It can be reduced to three material-specific parameters; the real-valued high-frequency limit of the permittivity $\varepsilon_\infty$ (F m$^{-1}$), the static conduction loss $\sigma_{dc}$ (S m$^{-1}$) and the (unit-less) electric susceptibility $\chi_r$. The frequency-dependent effective permittivity $\varepsilon_e(\omega)$ is given by:

$$\varepsilon_e(\omega) = \varepsilon_\infty + \varepsilon_0 \chi_r \left(\frac{\omega}{\omega_r}\right)\left[1 - \cot\left(\frac{n\pi}{2}\right)\right] - \frac{i\sigma_{dc}}{\omega} \tag{12}$$

where $\omega_r$ is an arbitrary frequency best chosen as the dominant frequency of the emitted antenna signal and *n* is a dimensionless empirical parameter that ranges from 0 to 1 and characterizes the magnitude of dielectric loss.

*3.4. Dipole discretization*

To ensure that the electric field generated by the source arrives approximately as a plane wave over each element, it is necessary to use an element discretization that is much smaller than the dominant wavelength of the emitted signal. Note that only the first intercepting plane of a fracture needs to be discretized so the discretization is 2D. To determine appropriate discretization criteria, we perform a synthetic Monte-Carlo simulation in which we generate 100 fracture realizations of random thickness, orientation, and length (fractures are square), filled with water of conductivity 0.1 S m$^{-1}$. The thicknesses were allowed to vary log-normally in the range of 0.1 mm to 10 cm while the orientation angles (0° to 60° in both dip and azimuth) and the fracture length (1 m to 10 m) were varied following a uniform distribution. The location of the midpoint of a fracture was randomly assigned to a maximum of 20 m away from the source location. The source and receiver were placed 3 m apart and we used a Gaussian distribution to generate the source dipole moment with a characteristic pulse corresponding to a dominant wavelength of 1 m, typical of a 100 MHz GPR antenna employed in crystalline rock.

To compare the gain in accuracy as a function of dipole discretization, we define a maximum discretization of 16 dipoles per (dominant) wavelength. This corresponds here to a dipole spacing of 6.67 cm in both tangential dimensions of a fracture. We compute the forward model response at the receiver location with the fine discretization and for successive coarsening using 8, 4, 2 and 1 dipoles per wavelength. For a given coarsening (index $c$), we compute the deviation between the finest discretization (index t), that we assume to be the true response, to the coarser discretization, as the Root Mean Square (RMS) difference of $N$ complex frequency amplitudes $A_i^{\{t,c\}}$ in the reflected response, given by:

$$\text{RMS}^c = \sqrt{\frac{1}{N}\sum_{i=1}^{N} \frac{|A_i^t - A_i^c|^2}{|A_i^t|^2}}. \tag{13}$$

We compute the RMS deviation at 500 linearly spaced frequencies, from 1 MHz to 1 GHz, and tabulate the maximum, mean, 25% and 75% quartiles for each coarsening level in Table 1. The statistics show that the error is very small for 8 dipoles per wavelength, while the value of 4 dipoles per wavelength appears also satisfactory with one outlier which gives a 15 % deviation and a mean error of 4 %. The outlier corresponds to a large fracture (8 m side length) with a small thickness (0.1 mm), large dip (37°) and azimuth (29°), and a center that is located only 3 m away from the antenna midpoint. It is often the case in practical applications that the energy close to the GPR system (early arrivals) is dominated by the direct wave and the experimental accuracy is not sufficient to infer information from reflections in the near-borehole region. A discretization of 4 dipoles per wavelength appears thus sufficient for most practical applications.

A possible limitation of the current implementation of our forward-modeling approach is that we do not consider secondary reflections from neighboring fracture elements. With a discretization of 4 dipoles per wavelength (which amounts to dipoles placed ~33 cm apart in this analysis) the magnitude of the antenna-emitted electric field arriving at the center of each element is several orders of magnitude higher than the field arriving from the other elements. Therefore, using only the antenna source for polarizing each element has a negligible effect on the scattered response of a collection of elements.

**Table 1:** Statistics of a Monte Carlo experiment with 100 fracture realizations. The measure used is the normalized root-mean-square difference in the (reflected) frequency response, at 500 frequencies spaced linearly from 1 MHz to 1 GHz, between the maximum (16 dipoles per wavelength) and coarser fracture discretizations.

| DIPOLE DENSITY | MAXIMUM | 75% QUARTILE | MEAN | 25% QUARTILE |
|---|---|---|---|---|
| 1 | 0.43 | 0.29 | 0.22 | 0.15 |
| 2 | 0.34 | 0.24 | 0.15 | 0.09 |
| 4 | 0.15 | 0.06 | 0.04 | 0.02 |
| 8 | 0.02 | 0.01 | 0.01 | 0.01 |

## 4. Results

In this section, we compare the results of our effective-dipole (ED) forward modeling scheme with a FDTD code and to data from a laboratory experiment described by Grégoire and Hollender (2004).

### 4.1. Experimental set-up of the synthetic experiment

The synthetic study offers a comparison of our effective-dipole forward modeling scheme with numerical simulations based on GPRMax3D, a 3D FDTD code that has been used extensively for modeling GPR responses (Giannopoulos, 2005). The experimental layout is shown in Fig. 2. The background matrix material is homogeneous with $\text{Re}\{\varepsilon_e\} = 5\,\varepsilon_0$ and $\sigma_{dc} = 0.001$ S m$^{-1}$, where $\sigma_{dc} = i\omega\,\text{Im}\{\varepsilon_e\}$. The reflector is a conductive water-filled fracture oriented in the $\hat{y}$-$\hat{z}$ plane with material properties $\text{Re}\{\varepsilon_e\} = 81\,\varepsilon_0$ and $\sigma_{dc} = 0.1$ S m$^{-1}$. The fracture (reflector) is square with a 2 m side length and 30 cm thickness. The thickness is chosen large enough to allow effective discretization with the FDTD code (2 cm node spacing in all 3 directions resulting in 15 nodes along the fracture thickness), but still within the thin-layer regime ($d = 0.3\,\lambda$) that is given by Bradford and Deeds (2006) as layer thicknesses less than 0.75 $\lambda$. For the FDTD code we used a spatial modeling domain consisting of a cube with an 8 m side length with absorbing boundaries (see bounding box in Fig. 2) to avoid boundary reflections.

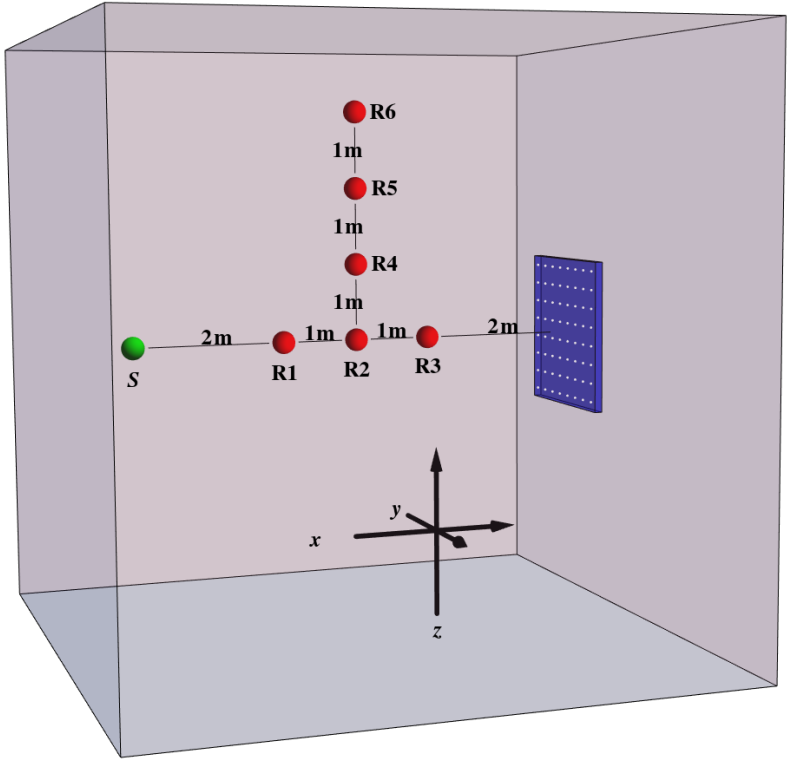

**Fig. 2.** Schematic of the synthetic experiment used to study propagation and reflection. The experiment coordinate system is shown with the axes in the bottom, receivers (red) are labelled R and the source (green) is labelled S. The square reflector is shown on the right (blue) with marks (white) indicating the dipole locations.

*4.2. Comparison of propagation results without fracture*

To compare propagation modeling results we use the horizontally placed receivers shown in Fig. 2. For the FDTD computation, we define a Ricker wavelet with central frequency of 100 MHz. The FDTD code implements a 'soft-source' by defining the current density over time at a given location in the grid to obtain a propagating Ricker wavelet. For our effective dipole computation we optimize the parameters of the source dipole moment distribution (described in section 3.2), starting with a Gaussian and using a local search algorithm to obtain a wavelet arriving at the first receiver, R1, that is similar, within 1% difference in amplitude and phase to the one obtained by the FDTD code. All subsequent results are normalized by the maximum amplitude of the propagating wavelet in R1, shown in Fig. 3a, and the same source parameters are used throughout section 4.3. The propagation results at each receiver location, R1, R2 and R3, are shown in Figs. 3a, 3b and 3c respectively. The results indicate that the two approaches produce very similar wavelets at all three receiver locations.

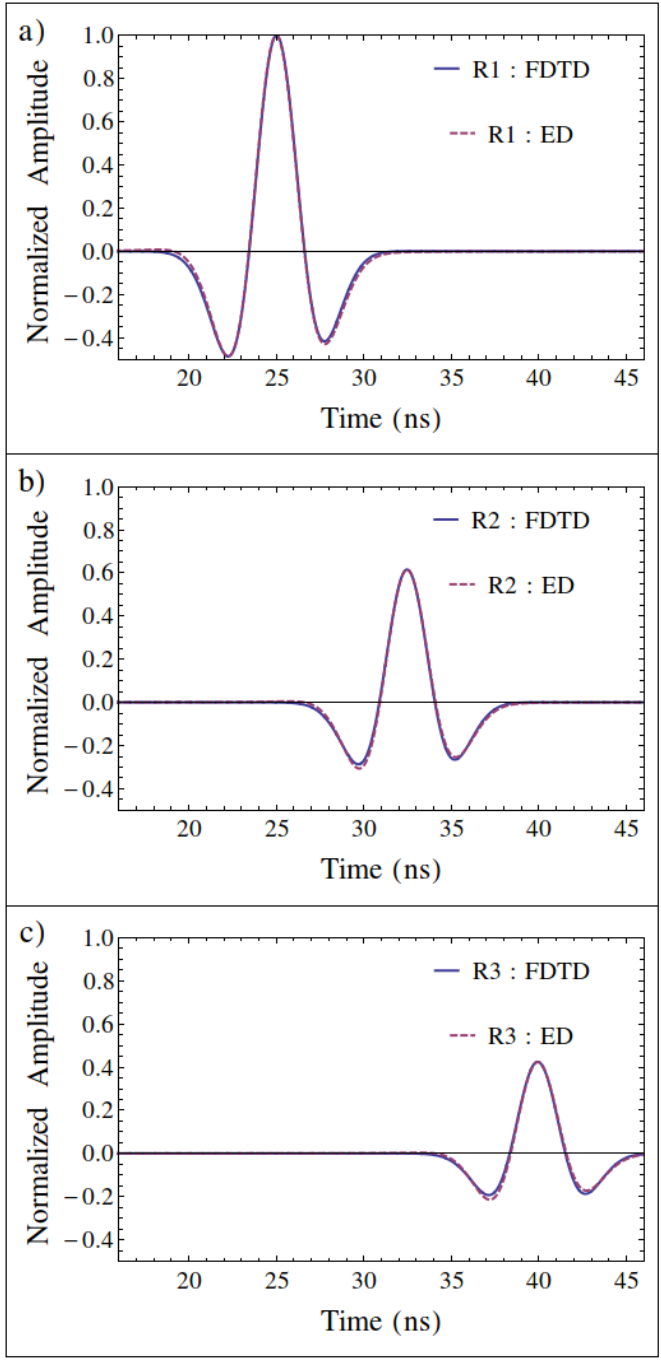

**Fig.3.** Comparison of propagation from our effective-dipole (ED) and FDTD forward models, for the three horizontal receiver locations R1, R2 and R3 in Fig. 2.

*4.3. Comparison of reflection results*

To compare the differences in the reflected signal between our method and FDTD, we use both the horizontally placed receivers (R1, R2 and R3) and the vertically placed receivers (R4, R5 and R6) in Fig. 2. The results are presented in Figs. 4 and 5 respectively. The source pulse is generated using the same parameters and normalization as in the previous section. For all reflections, we see very good agreement between the early arriving energy, in both amplitude and phase, while the later arriving energy shows minor mismatches between the two methods.

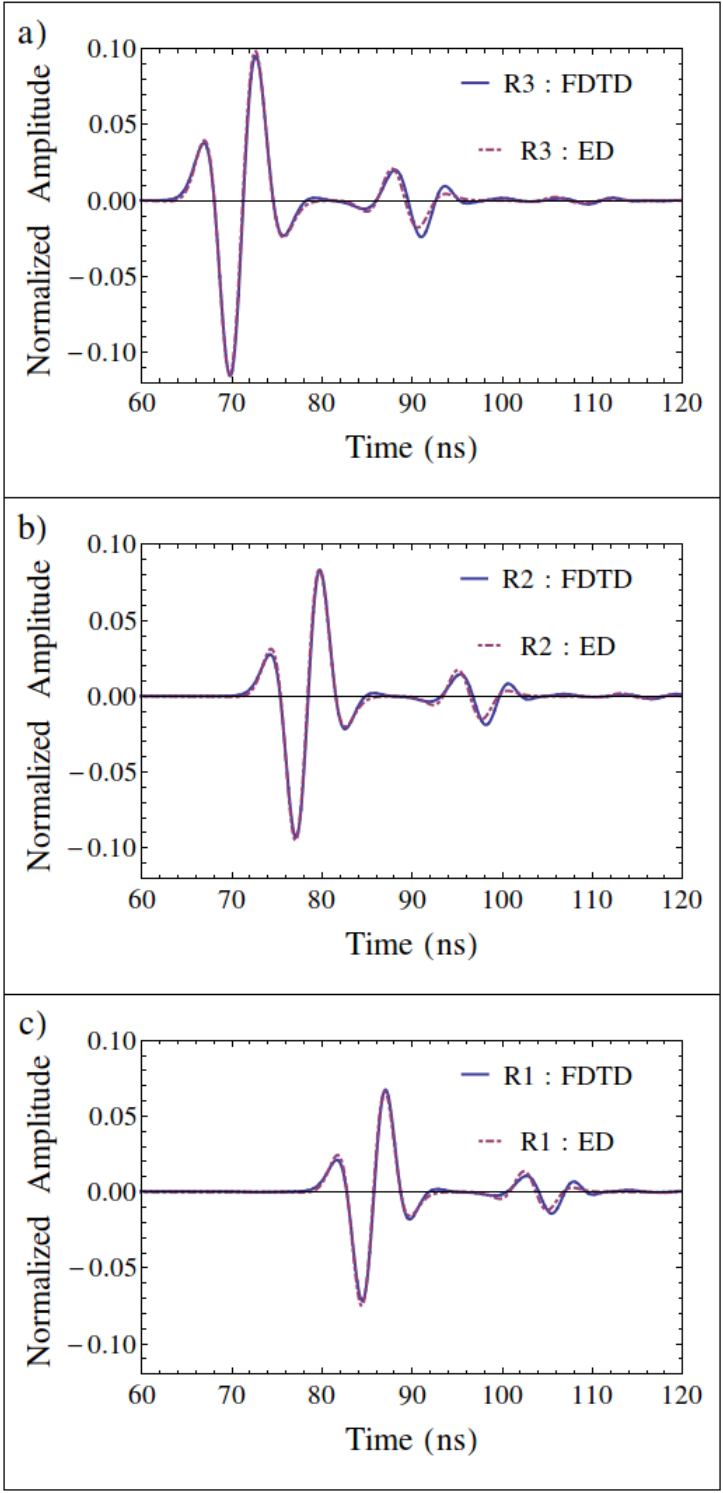

**Fig. 4.** Comparison of reflections from our effective-dipole (ED) and FDTD forward models, for the three horizontal receiver locations R1, R2 and R3 in Fig. 2.

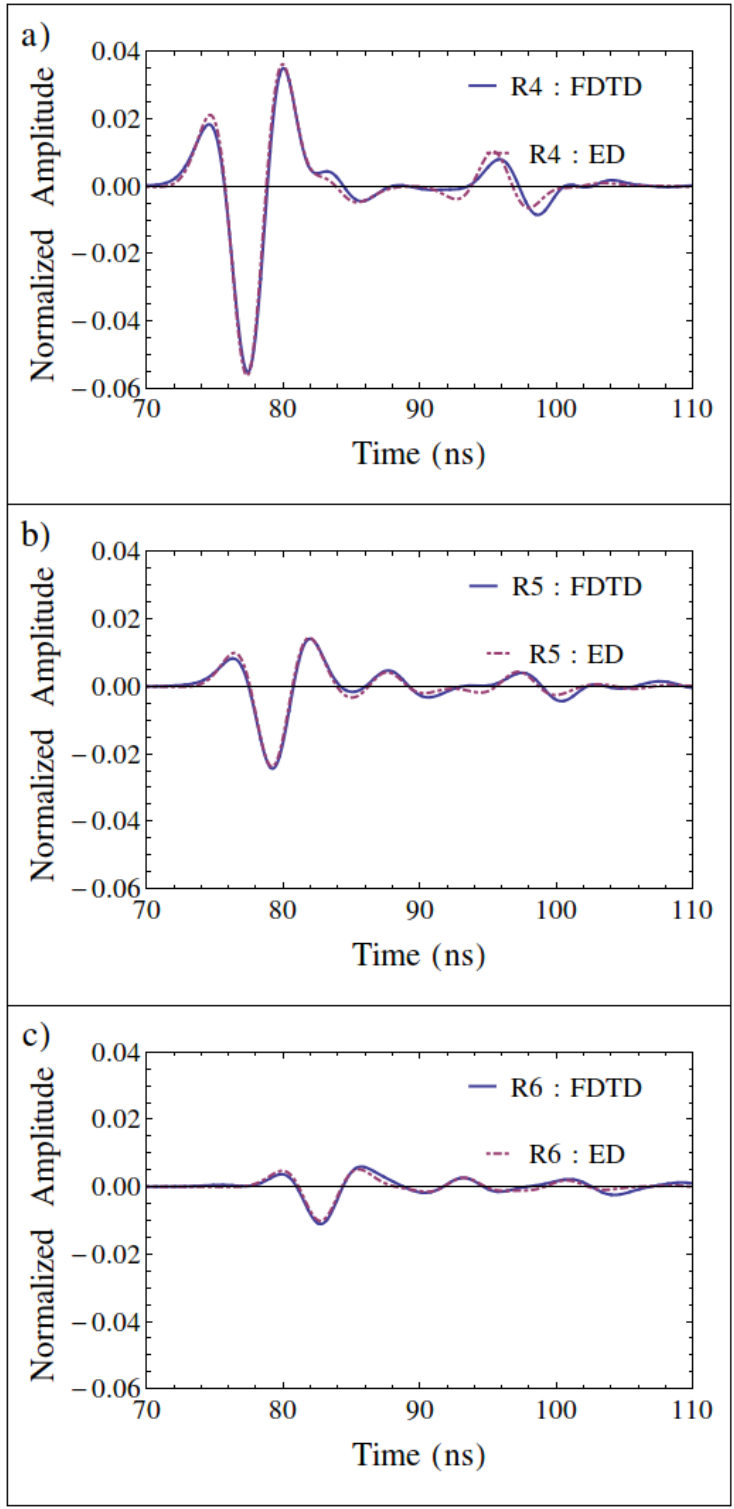

**Fig. 5.** Comparison of reflections from our effective-dipole (ED) and FDTD forward models, for the three vertical receiver locations R4, R5 and R6 in Fig. 2.

*4.4. Comparison to laboratory results*

In this section, we compare the results of our effective-dipole forward modeling scheme to the laboratory results presented by Grégoire and Hollender (2004). In these experiments (see Fig. 6), two granitic blocks are held apart by 5 and 2.5 mm, while the separation between the blocks is filled with materials of varying properties. The electric properties of the filling materials are measured in the laboratory using a dielectric probe kit, and a 900 MHz bi-static GPR antenna is used to measure the reflection arising from the thin layer. It was not possible to simulate this experiment with the FDTD code because of memory limitations in defining a sub-millimeter discretization in a meter-scale 3D domain.

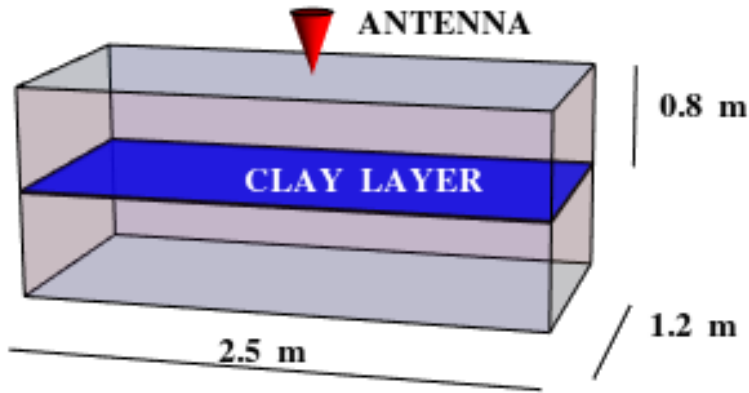

**Fig. 6.** Schematic of the laboratory setup by Grégoire and Hollender (2004). The rectangular (blue) sheet indicates the location of the reflective clay layer and the (red) cone indicates the location of the GPR antenna while the bounding cuboid (gray) represents the two granite blocks surrounding the clay layer.

As a first comparison between our effective-dipole simulation and the laboratory data, we use a Gaussian pulse with a central frequency around 900 MHz. We allow for frequency-dependent attenuation and dispersion using the Jonscher parametrization given in Eq. (12). Grégoire and Hollender (2004) give a set of reduced parameters that they use in an alternative formulation of the Jonscher parametrization. From these parameters it is possible to derive the original Jonscher parameters (see Table 2). Following Grégoire and Hollender (2004), we present the laboratory and simulated data (Fig. 7) normalized to the reflection from the 5 mm clay layer. The observed amplitude difference between the 5 mm and 2.5 mm saturated clay layer is reproduced well, but the wavelet shape and duration is poorly reproduced by using the Gaussian antenna-pulse. Figure 8 shows a comparison between the laboratory-measured and simulated (reflected) frequency amplitudes from the 5 mm layer.

**Table 2:** Jonscher parameters used, to account for frequency dependent attenuation and dispersion, in simulating the laboratory results of Grégoire and Hollender, (2004). The first three parameters are unit-less.

| MATERIAL (Units) | $n$ | $\chi$ | $\varepsilon_\infty/\varepsilon_0$ | $2\pi\omega_c$ (MHz) | $\sigma_{dc}$ (S m$^{-1}$) |
|---|---|---|---|---|---|
| Granite | 0.93 | 0.7 | 5.3 | 800 | 0.003 |
| Saturated Clay | 0.69 | 14. | 47. | 800 | 0.68 |

We now investigate to what extent the agreement between the forward simulations and the laboratory data can be improved by allowing the shape of the antenna pulse to vary according to Eq. (11). For the 5 mm experiment, we search for the pulse that gives the smallest RMS, as defined in Eq. (13), between the laboratory-measured and modeled reflected energy spectrum, at 10 linearly spaced frequencies. We use a local search algorithm to compute the optimal pulse and converge to a solution with 7% RMS error, compared to the initial RMS of 30% that was obtained by using the Gaussian pulse as a source. In Fig. 8 we show the reflected frequency spectrum, obtained by applying the optimal pulse through Eq. (11), along with the frequencies used for the optimization. The optimal antenna pulse generates a signal with a significantly lower frequency content than the purely Gaussian pulse.

Our simulation results (Fig. 7) show that optimizing a generalized Gamma distribution to model the antenna-pulse shape can help to reproduce the pulse width and location of maxima and minima, as well as complex patterns in the signal such as the wide peak appearing after 3 ns. The amplitude and phase in the reflected signals seem to coincide, for the most part, except for the initial and final peaks that are not exactly reproduced in our simulations.

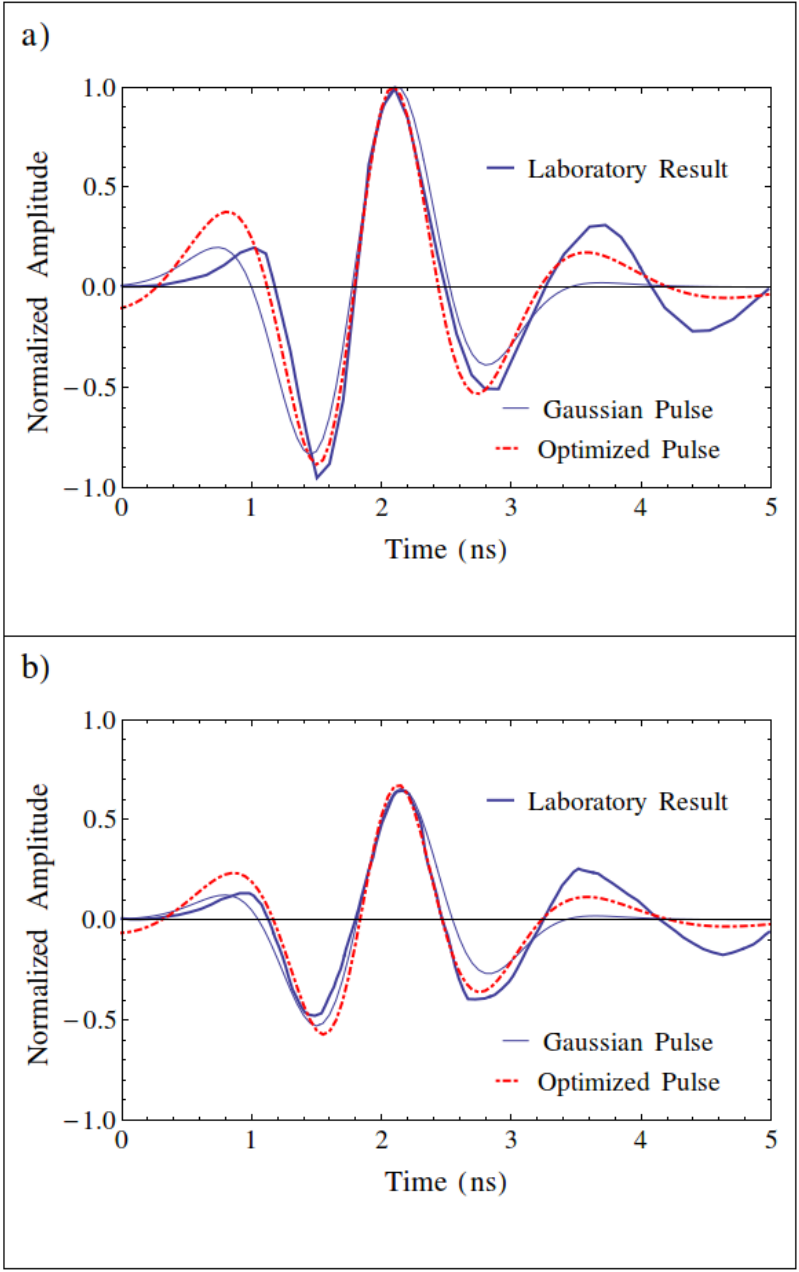

**Fig. 7.** Comparison of the reflected electric field acquired from laboratory data of Grégoire and Hollender (2004) and computed using the effective-dipole forward model with a Gaussian pulse and an optimized pulse as the antenna-source. Results are shown for (a) a 5 mm layer and (b) a 2.5 mm layer.

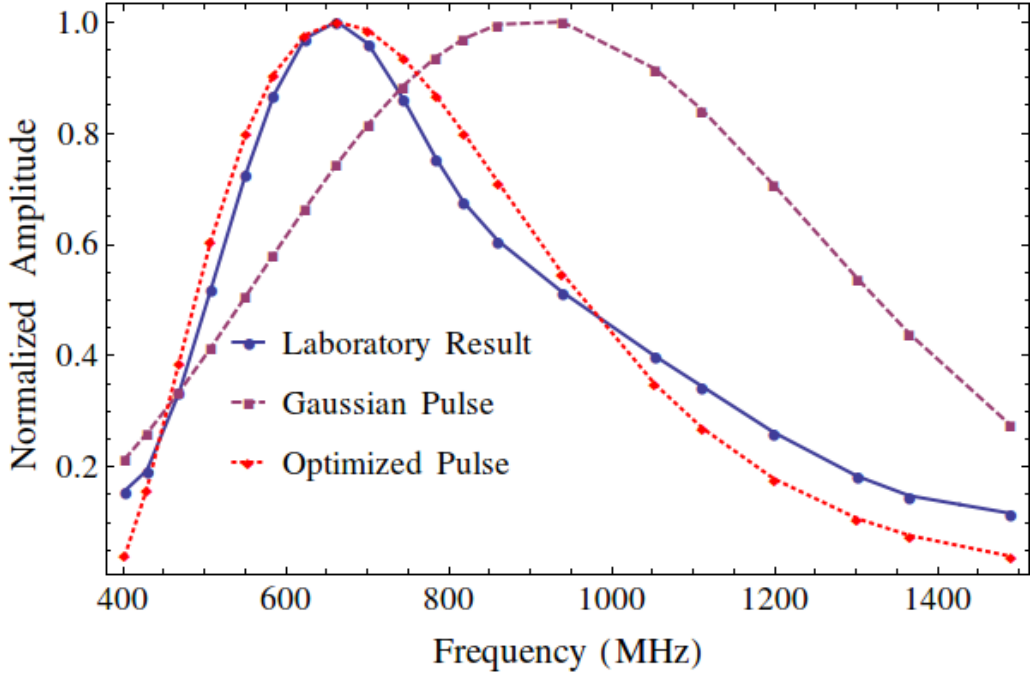

**Fig. 8.** Comparison between the reflected frequency spectra of the Grégoire and Hollender (2004) laboratory experiment (5 mm layer) using a Gaussian pulse, centered around 900 MHz, and an optimized pulse. The frequencies used for the optimization of the pulse are shown with markers on the plots and the parameters used to generate the optimal antenna pulse in Eq. (11) are $\alpha = 4$; $\beta = 0.05$; $\gamma = 0.79$; $\mu = 0.38$.

## 5. Discussion

The comparison between our effective-dipole method and FDTD simulations provides many illuminating results. Primarily, propagation of the electric field (Fig. 3) compares very well both in terms of attenuation and dispersion of the wavelet. Additional tests (not shown here) demonstrate significant numerical grid dispersion (e.g., when using a slightly coarser discretization of 3 cm) in the FDTD results, which is not a problem in our approach as analytic closed-form solutions are used in all calculations.

The reflected wavelets (Figs. 4 and 5) show some subtle differences between the two methods. For both horizontally and vertically placed receivers the results agree well between the FDTD and our effective-dipole formulation in both amplitude and phase. At closer inspection, one can see minor discrepancies in the later arrivals of the reflected energy. We have performed tests with higher conductivities of the water-filled fracture where the later arrivals are not visible, and instead only the first four peaks are prominent and agree well (as shown here) between the two methods. We postulate that these later arrivals are a result of the infinite internal reflections within the fracture, which are calculated in fundamentally different ways for the two methods. In our effective-dipole formulation we use analytical solutions from optics to account for the internal reflections while the FDTD code accounts for these iteratively. The latter would only approach the exact limit in the case of infinitely fine temporal sampling.

While numerical dispersion can be counteracted using finer discretizations, and boundary reflections can be minimized using a larger domain and/or more effective boundary conditions, it is often computationally demanding to apply efficient discretization schemes in an effective manner. This is especially evident when studying fractures with thicknesses at the millimeter scale in a domain of several meters, let alone tenths of meters as in most field applications. Recent FDTD work has been focused on using sub-grid discretizations to model the interaction of EM waves with very small layers (Diamanti and Giannopoulos, 2009), but the problem of efficiently discretizing irregular geometries (e.g., highly dipping fractures) remains and the computational constraints are still high for 3D implementations. Our approach does not suffer from these drawbacks, and layers of arbitrarily small thickness can be considered at any distance from the source. Furthermore, there are no boundary effects in our formulation.

The comparison to the laboratory data, shown in Fig. 7, suggests that the effective-dipole forward model is physically sound, since the phase delays and, especially, the amplitudes of the measured and modeled wavelets match well. Furthermore, using the generalized Gamma distribution to define the dipole moment of the source allows us to model an effective signal that accounts for uncertainties in the emitted signal in a satisfactory manner. The optimized pulse generates a signal with significantly lower frequency content than the equivalent Gaussian pulse with a peak at the antenna dominant frequency, as is expected in practice for GPR applications.

For our effective-dipole model, the forward model has to be run individually for each source-receiver combination, but the approach is fully parallelizable and can be easily implemented using parallel computing. Computationally, the effective-dipole model provides a faster alternative to traditional numerical approaches and can decrease computation times by several orders of magnitude. In the synthetic experiment, the FDTD simulations takes approximately two hours while the effective-dipole model takes only 30 seconds to compute the response for each source-receiver pair, totaling 3 minutes for the whole experiment (2.9 GHz CPU with 7.5 GB RAM PC running Ubuntu). Typical discretizations for practical applications may be on the order of 1000 dipoles, for which one frequency component can be computed in approximately 0.5 second on a standard PC.

**6. Conclusions**

We present a forward model that describes the propagation of electromagnetic waves through a dense homogeneous dielectric medium and scattering at thin planar layers. We discretize the layers into elements that respond as polarized dipoles, modulated by the effective reflection coefficients of thin layers. We account for frequency-dependent electrical properties of the media through the Jonscher formulation, and model uncertainties in the emitted signal by using a generalized Gamma distribution as a current source. Our model compares well with finite-difference time-domain (FDTD) computations and does not suffer from numerical inaccuracies or boundary effects. Compared to FDTD, we are able to introduce reflectors of arbitrary size, thickness, material filling and orientation without compromising accuracy. We are also able to simulate laboratory data that we are not able to simulate with the FDTD approach. Optimizing the pulse shape through the generalized Gamma distribution further improved the agreement with the laboratory data. Computation times are orders of magnitudes smaller than FDTD and the approach is easily parallelizable. This forward modeling approach will soon be coupled with flow and transport simulations in discrete fracture networks to infer transport behavior at experimental hydrological field sites.

**Acknowledgements**
This research was supported by the Swiss National Science Foundation under grant 200021-146602. We would like to thank George Tsoflias and Jan van der Kruk for their thorough feedback during the review process, as well as Steward Greenhalgh for his extensive comments that helped to improve this manuscript.

National Research Council (US), 1996. Committee on Fracture Characterization, & Fluid Flow. Rock fractures and fluid flow: contemporary understanding and applications. Natl Academy Pr.

Neuman, S.P., 2005. Trends, prospects and challenges in quantifying flow and transport through fractured rocks. Hydrogeol. J. 13, 124–147.

Olsson, O., Falk, L., Forslund, O., Lundmark, L., Sandberg, E., 1992. Borehole radar applied to the characterization of hydraulically conductive fracture zones in crystalline rock. Geophysical Prospecting 40, 109–142.

Purcell, E.M., Pennypacker, C.R., 1973. Scattering and absorption of light by nonspherical dielectric grains. Astrophys. J. 186, 705–714.

Purcell, E.M., Smith, A.S., 1986. Electricity and magnetism. Berkeley Physics Course (Especially Ch. 10), Vol. 3, 2nd Edition, McGraw-Hill, New York

Russakoff, G., 1970. A derivation of the macroscopic Maxwell equations. Am. J. Phys. 38, 1188–1195.

Sambuelli, L., Calzoni, C., 2010. Estimation of thin fracture aperture in a marble block by GPR sounding. Boll. Geofis. Teor. Ed Appl. 51, 239–252.

Sassen, D.S., Everett, M.E., 2009. 3D polarimetric GPR coherency attributes and full-waveform inversion of transmission data for characterizing fractured rock. Geophysics 74, J23–J34.

Stacy, E.W., 1962. A generalization of the gamma distribution. Ann. Math. Stat. 1187–1192.

Talley, J., Baker, G.S., Becker, M.W., Beyrle, N., 2005. Four dimensional mapping of tracer channelization in subhorizontal bedrock fractures using surface ground penetrating radar. Geophys. Res. Lett. 32.4

Tsoflias, G.P., Becker, M.W., 2008. Ground-penetrating-radar response to fracture-fluid salinity: Why lower frequencies are favorable for resolving salinity changes. Geophysics 73, J25–J30.

Tsoflias, G.P., Halihan, T., Sharp, J.M., 2001. Monitoring pumping test response in a fractured aquifer using ground-penetrating radar. Water Resour. Res. 37.5, 1221–1229.

Tsoflias, G.P., Hoch, A., 2006. Investigating multi-polarization GPR wave transmission through thin layers: Implications for vertical fracture characterization. Geophys. Res. Lett. 33.2031

# FIGURES AND TABLES